\documentclass[aps,prl,floatfix,twocolumn,footinbib]{revtex4}
\usepackage{epsfig}
\usepackage{amsmath, amssymb, amsthm, amsfonts}
\usepackage{dsfont, bm}
\usepackage{braket}
\usepackage[english]{babel}

\newcommand{\ii}{\operatorname{i}}
\renewcommand{\vec}[1]{#1}
\newcommand{\mat}[1]{#1}
\newcommand{\op}[1]{\hat{#1}}

\newcommand{\id}{\mathds{1}}
\newcommand{\identityop}{\mathds{1}}
\newcommand{\zeroop}{\mathds{O}}
\newcommand{\defemph}[1]{\emph{#1}}
\newcommand{\fourmatrix}[4]{\begin{pmatrix} #1 & #2 \\ #3 & #4 \end{pmatrix}} 
\newcommand{\transpose}[1]{#1 ^{\operatorname T}}
\renewcommand{\det}[1]{\operatorname{det} \left[ #1 \right]}
\newcommand{\half}{\frac{1}{2}}
\theoremstyle{plain}
\newtheorem{Satz}{Satz}[]
\newtheoremstyle{defn} 
{} 
{6mm} 
{\itshape} 
{} 
{\bfseries} 
{:} 
{1em} 
{} 
\theoremstyle{defn}
\newtheorem{proposition}[Satz]{Proposition}

\bibliography{references}
\begin{document}
\title{Separability Criterion for One-Sided Gaussian Channels}
\author{Jason Hoelscher-Obermaier}
\author{Peter van Loock}\email{peter.vanloock@mpl.mpg.de}
\affiliation{Optical Quantum Information Theory Group, Max Planck Institute for the Science of Light and Institute of Theoretical Physics I, Universit\"at Erlangen-N\"urnberg, Staudtstr. 7/B2, 91058 Erlangen, Germany}

\begin{abstract}
We show that the following nontrivial necessary precondition for an entanglement evolution equation for pure Gaussian states under one-sided Gaussian channels holds. Suppose a Gaussian quantum channel acts on one mode of a pure entangled multi-mode Gaussian input state. Then, for a fixed channel, either all output states are entangled or none of them are. In other words, if the input state is Gaussian, pure and entangled, the separability after a one-sided Gaussian quantum channel does not depend on the input state, but only on the channel. Furthermore, a simple linear-algebraic separability criterion allows to decide whether a given channel destroys the entanglement of pure entangled input states or leaves them entangled.
\end{abstract}
\maketitle

\section{Motivation}
\label{sec:motivation}
A large fraction of the known quantum information protocols (QIPs) --- including such important protocols as quantum key distribution \cite{curty04} and measurement-based quantum computation \cite{raussendorf01} --- relies on entangled states. Unfortunately, entanglement is a quite fragile resource and unwanted interactions with the environment tend to destroy it rapidly. In order to assess the feasibility of many entanglement-based QIPs, it is therefore important to know how entanglement evolves in time.

A quantum system in state $\rho$ which interacts with an environment in state $\rho_E$ is in general subject to a non-unitary time evolution $T$
\begin{equation}
T:\quad \rho \mapsto T(\rho)= \mathrm{tr}_E \left[ U (\rho \otimes \rho_E) U^\dagger \right] , \label{eq:general_quantum_channel}\end{equation}
 where $U$ is the time evolution of both the system and its environment. $T$ is a completely positive, trace-preserving map from the set of density operators on a certain Hilbert space $\mathcal H$ into itself, a so-called \defemph{quantum channel}.

The standard way to calculate the evolution of entanglement under a given quantum channel is to directly calculate the evolution of the quantum state in question according to \eqref{eq:general_quantum_channel} and then calculate the entanglement of the output state $ T(\rho)$. In particular, the calculation has to be performed separately for every input state. In the general case of arbitrary quantum channels and arbitrary (pure as well as mixed) input states, this is the only viable way.

Fortunately, we do not have to consider the most general quantum channel since typical noise effects can be described as \defemph{local quantum channels}, that is, they can be considered as acting uncorrelated on the different parts of the quantum system. It is therefore practically relevant to characterize the evolution of entanglement under local quantum channels \footnote{Local quantum channels also play a very important role in quantum communication because they include the lossy and noisy fiber channels which come up in the context of entanglement distribution.}. As a special case of local quantum channels, this paper considers \defemph{one-sided quantum channels} which act only on a single qubit or a single mode of an entangled state.

\section{Qubit Entanglement Evolution Equation}
\label{sec:qubit_entanglement_evolution}
In the qubit case, the restriction to one-sided quantum channels is known to simplify the calculation of the entanglement evolution drastically. More specifically, consider an arbitrary one-sided quantum channel $\mathdollar$ which acts on one qubit of an initially pure two-qubit state $\ket \chi$ (hence the output state is $(\id\otimes\mathdollar)[\ket{\chi}\bra{\chi}]$) and consider the evolution of the concurrence $C$ under this channel \footnote{The use of the concurrence is essential because the proof of \eqref{eq:konrad_ent_evolution} relies on the analytic formula for the concurrence of arbitrary (pure and mixed) two-qubit states~\cite{wootters98}.}.
Konrad et al.~\cite{konrad08} have shown that, for all pure two-qubit input states $\ket\chi$, the concurrence of the output state is given by
\begin{equation}
 \label{eq:konrad_ent_evolution}
C \left\lbrace ( \id \otimes \mathdollar )[\ket{\chi}\bra{\chi}] \right\rbrace =
C \left\lbrace ( \id \otimes \mathdollar )[\ket{\Phi}\bra{\Phi}] \right\rbrace \, \cdot \, C \left\lbrace \ket{\chi}\bra{\chi} \right\rbrace ,
\end{equation}
where $\ket\Phi$ is one of the Bell states. Hence, up to a prefactor given by the entanglement of the input state, the entanglement (as quantified by the concurrence) evolves always exactly like the entanglement of the Bell states.

In particular, this implies that a given one-sided quantum channel will either disentangle all initially entangled pure two-qubit states or leave all of them entangled. For if the initial state $\ket \chi$ is entangled, we have $C \left\lbrace \ket{\chi}\bra{\chi} \right\rbrace \neq 0$ and therefore, according to \eqref{eq:konrad_ent_evolution}, the entanglement of the output state is zero if and only if $ C \left\lbrace ( \id \otimes \mathdollar )[\ket{\Phi}\bra{\Phi}] \right\rbrace = 0 $ --- independent of the state $\ket \chi$.

\section{Gaussian Entanglement Evolution}
\label{sec:qumode_entanglement_evolution}
Equation \eqref{eq:konrad_ent_evolution} is relevant because qubits are ubiquitous in quantum information. But in many quantum-optical implementations of QIPs, continuous variables are preferred over qubits because of practical advantages \footnote{For example, continuous variable implementations are usually deterministic while qubit protocols are often probabilistic. See, for example, \cite{braunstein05} for a recent review of continuous variable quantum information processing.}.
In continuous variable quantum information processing, Gaussian states are widely used and typical errors are described by Gaussian quantum channels. Therefore, an evolution equation for the entanglement of Gaussian states under one-sided Gaussian channels similar to \eqref{eq:konrad_ent_evolution} would be highly relevant. But, an equation of the form~\eqref{eq:konrad_ent_evolution} in the Gaussian regime would require that --- just as in the qubit case --- the separability after the one-sided Gaussian channel depends only on the channel and not on the input state. Before this fact is proved, some basic concepts of the theory of Gaussian states have to be introduced.

\subsection{Gaussian Channels and Covariance Matrices}
\label{sec:gauss_channels}
Consider a mode of the electro-magnetic field with annihilation and creation operators $\op a, \op a ^\dagger$. The \defemph{quadratures} $\op q , \op p $ are defined by
\begin{equation}
\label{eq:quadratures}
 \op q = \frac{1}{\sqrt 2} \left( \op a + \op a ^\dagger \right), \quad \op p = \frac{1}{\sqrt 2 \ii} \left( \op a - \op a ^\dagger \right) .
\end{equation}
Every quantum state of the field corresponds uniquely to a Wigner function which can be considered as a quasi-probability distribution over the possible measurement results for the quadratures. \defemph{Gaussian states} are states whose Wigner function is a Gaussian distribution. In particular, they are completely characterized by the expectation values and variances of the quadrature operators. The expectation values can be set to zero by local operations and are therefore irrelevant for the entanglement of the state. %
All the interesting information is contained in the variances which are conveniently summarized in the \defemph{covariance matrix} (CM). The CM of a two-mode state $\rho_{AB}$ on modes $A$ and $B$ is a real symmetric $4 \times 4$ matrix ${\mat\sigma_{AB}}$ with entries
\begin{equation}
 {\left( \mat\sigma_{AB} \right)}_{ij} = \half \braket{ \op R_i \op R_j + \op R_j \op R_i} - \braket{\op R_i}\braket{\op R_j} ,
\end{equation}
where $\vec{\op R} = (\op q_A, \op p_A, \op q_B, \op p_B)$ is the vector containing the quadratures and $\braket{\op X}$ is the expectation value of the operator $\op X$ in the state $\rho_{AB}$. Note that not every real symmetric $4 \times 4$ matrix is the CM of a quantum state since the variances have to fulfill the Robertson-Schr\"odinger inequality. If we write $\mat \sigma_{AB}$ in block form \begin{equation}
\mat \sigma_{AB} = \fourmatrix{\mat \alpha}{\mat \gamma}{\transpose{\mat\gamma} }{\mat\beta}
\end{equation}
(with real $2\times2$ matrices $\mat\alpha,\mat\beta, \mat \gamma$), the Robertson-Schr\"odinger inequality is fulfilled exactly if (see \cite{ferraro05})
\begin{equation}
 \label{eq:ferraro_physicality_condition}
\det{\mat\alpha}+\det{\mat\beta} + 2\det{\mat\gamma} \leq 4 \det{\mat\sigma_{AB}} + 1/4 .
\end{equation}
For future reference we also note that the separability of $\mat \sigma_{AB}$ is equivalent to
\begin{equation}
 \label{eq:ferraro_separability_condition}
\det{\mat\alpha}+\det{\mat\beta} - 2\det{\mat\gamma} \leq 4 \det{\mat\sigma_{AB}} + 1/4 .
\end{equation}
The most important example of an entangled two-mode Gaussian state is the two-mode squeezed state with CM ($\identityop_n$ denotes the $n\times n $ identity matrix)
\begin{align}
& \mat\gamma_{\mathrm{TMSS,r}} = \half \fourmatrix{\cosh ( r ) \identityop_2}{\sinh ( r ) \mat \sigma_z}{\sinh ( r ) \mat \sigma_z}{\cosh ( r ) \identityop_2} \nonumber \\
& \mat \sigma_z = \fourmatrix{1}{0}{0}{-1}, \; r>0 .
\end{align}

A \defemph{Gaussian quantum channel} is defined as a quantum channel which maps Gaussian states onto Gaussian states. To calculate the evolution of the entanglement, we only need to consider the action of the channel on the CM. For future convenience, let us define the abbreviation $\mat A \left[ \mat B \right] \equiv \mat A \mat B \transpose{\mat A}$ if $\mat A$ and $\mat B$ are matrices. If a Gaussian channel $\mathcal G$ acts only on mode $B$ of a two-mode Gaussian state, the initial CM $\sigma_{AB}$ evolves as (see \cite{eisert05})
\begin{equation}
\label{eq:one-sided_channel_output_CM}
\mat\sigma_{AB} \overset{\identityop\otimes\mathcal G}{\longrightarrow} (\mathds 1 \oplus \mat f) [\mat\sigma_{AB}] + (\zeroop \oplus \mat g) ,
\end{equation}
where $\mat f $ and $\mat g$ are two real $2\times 2$ matrices which completely characterize the channel $\mathcal G$, $\zeroop$ represents the zero matrix, and $\oplus$ denotes the matrix direct sum. Since the $\mat g $-contribution is independent of the state $\mat\sigma_{AB}$, it can be thought of as a noise term. Note that not every pair of matrices $\mat f , \mat g$ describes a Gaussian quantum channel since the output state must again respect the Robertson-Schr\"odinger inequality as embodied by \eqref{eq:ferraro_physicality_condition}. Channel matrices $\mat f , \mat g$ which describe a genuine Gaussian channel fulfill (see \cite{eisert05})
\begin{equation}
4 \det{\mat g} \geq (\det{\mat f} - 1 )^2 .
\label{eq:one-sided-ch_matrix_condition}
\end{equation}

\subsection{Separability Criterion for Gaussian Channels}
\label{sec:no_partially_disent_channels}
\begin{proposition}
\label{th:no_partially_disent_channels}
If a one-sided Gaussian quantum channel $\mathcal G$, which is characterized by channel matrices $\mat f$ and $\mat g$, acts on mode $B$ of a pure and entangled two-mode Gaussian state with covariance matrix $\mat\sigma_{AB}$, then the output state $ {(\mathds 1 \oplus \mat f) [\mat\sigma_{AB}] + (\zeroop \oplus \mat g)} $ is separable if and only if
\begin{equation}
\label{eq:channel_sep_condition_ineq}
4 \det{\mat g} \geq (\det{ \mat f} + 1)^2 .
\end{equation}
The same condition \eqref{eq:channel_sep_condition_ineq} is also equivalent to the separability of the output state if the channel acts on a single mode of a Gaussian state on more than two modes.

\end{proposition}
This means that, for initially pure and entangled states, the separability of the output state depends only on the properties of the channel and not on the properties of the initial state. Accordingly, every one-sided Gaussian channel can easily be classified as ``disentangling'' or ``non-disentangling'' via the separability condition \eqref{eq:channel_sep_condition_ineq}. Note that this is quite counterintuitive: For any two Gaussian states $X$ and $Y$ --- no matter how slightly entangled $X$ and how strongly entangled $Y$ --- there is no one-sided Gaussian channel that destroys the entanglement of $X$ and leaves some of the entanglement of $Y$ intact \footnote{This is particularly counterintuitive since the entanglement of Gaussian states is unbounded. Hence, the entanglement of the state $Y$ can be arbitrarily high.}.

Holevo \cite{holevo08} provided general conditions under which Gaussian channels disentangle all input states, which are in agreement with \eqref{eq:channel_sep_condition_ineq}. His results, however, do not imply that any Gaussian channel which leaves some pure entangled Gaussian input state entangled must leave all such states entangled, as required for the existence of a Gaussian entanglement evolution equation.

Note also the close similarity of the separability condition \eqref{eq:channel_sep_condition_ineq} and the physicality condition \eqref{eq:one-sided-ch_matrix_condition} for the channel matrices. %
For any two real matrices $\mat f, \mat g$, there is a quantum channel such that $\mat f, \mat g$ describe its action on CMs according to \eqref{eq:one-sided_channel_output_CM} if the determinant of the noise matrix $\mat g$ is large enough, i.e. if $4 \det{\mat g} \geq (\det{\mat f} - 1)^2$. If $\mat g$ also fulfills $4 \det{\mat g} \geq (\det{\mat f} + 1)^2$, this channel will only output separable states.

\begin{proof}[Proof of Proposition 1]
Every pure, entangled two-mode Gaussian state $\mat\sigma_{AB}$ is related to a two-mode squeezed state $ \mat \gamma_{\mathrm{TMSS,r}}$ via local unitary operations which act on the CM as symplectic matrices $\mat S_A$ and $\mat S_B$ \cite{simon00}
\begin{equation}
\label{eq:relation_to_TMSS}
 \mat\sigma_{AB} = (\mat S_A\oplus \mat S_B) [\mat\gamma_{\mathrm{TMSS,r}}] ,
\end{equation}
where again the abbreviation $\mat A[\mat B] \equiv \mat A \mat B \transpose{\mat A}$ has been used.
The transformation $\mat S_A$ of mode $A$ commutes with the channel which acts on mode $B$ and can be undone after the channel since local unitary operations do not change the entanglement of the output state. Therefore, using \eqref{eq:relation_to_TMSS}, the separability of the output state $ {(\mathds 1 \oplus \mat f) [\mat\sigma_{AB}] + (\zeroop \oplus \mat g)} $ is equivalent to the separability of the state
\begin{align}
\label{eq:output_state_simplified}
  (\mathds 1 \oplus \mat f \mat S_B) [\mat\gamma_{\mathrm{TMSS,r}}] + (\zeroop \oplus \mat g)
&=
\fourmatrix{\mat C}{\mat S (\mat f \mat S_B )^T}{\mat f \mat S_B \mat S}{\mat f \mat S_B \left[\mat C\right] + \mat g} \nonumber\\
&\equiv
\fourmatrix{\mat \alpha}{\mat\gamma}{\transpose{\mat\gamma}}{\mat\beta} ,
\end{align}
where $\mat C\equiv 1/2 \cosh(r) \identityop_2$ and $ \mat S \equiv 1/2 \sinh(r) \mat\sigma_z$. According to \eqref{eq:ferraro_separability_condition}, the separability of \eqref{eq:output_state_simplified} is equivalent to
\begin{align}
 \det{\mat C} + \det{\mat{f S_B \left[ C\right] + g}} - 2 \det{\mat{f S_B S}} \nonumber \\
\leq 4 \det{ (\mathds 1 \oplus \mat f \mat S_B) [\gamma_{\mathrm{TMSS,r}}] + (\zeroop \oplus \mat g) } + 1/4 .
\label{eq:sep_inequality_with_dets_pure}
\end{align}

Now there are two cases. %
Assume first that $\det{\mat f} = 0$. %
In this case, the channel separability condition \eqref{eq:channel_sep_condition_ineq} is automatically fulfilled because any Gaussian channel also fulfills \eqref{eq:one-sided-ch_matrix_condition}. Therefore, it has to be shown that, in this case, the output state is indeed separable, as implied by \eqref{eq:channel_sep_condition_ineq} together with \eqref{eq:one-sided-ch_matrix_condition}.%

For $\det{\mat f} = 0$, we have $\det{\mat \gamma} = \det{\mat{f S_B S}}= \det{\mat f} \det{\mat{S_B S}} =0$ and the separability condition \eqref{eq:ferraro_separability_condition} reduces to
\begin{align}
\label{eq:detf0_physicality}
 \det{\mat\alpha} + \det{\mat\beta} - 2 \det{\mat\gamma} &\overset{\det{ \mat\gamma}=0}{=} \det{ \mat\alpha} + \det{ \mat\beta} \nonumber\\
 &\leq 4 \det{ \mat\sigma} + 1/4
\end{align}
But, under the same conditions (for $\det{\mat f}=0$), the separability condition \eqref{eq:detf0_physicality} is implied by condition \eqref{eq:ferraro_physicality_condition} which expresses the physicality of the output CM:
\begin{align}
\det{\mat\alpha} + \det{\mat \beta} + 2 \det{ \mat\gamma} &\overset{\det{ \mat\gamma}=0}{=} \det{ \mat\alpha} + \det{ \mat\beta} \nonumber\\
&\leq 4 \det{ \mat\sigma} + 1/4 .
\end{align}
Therefore, if $\det{\mat f} = 0$, the output state is indeed separable.

Assume now that $\det{\mat f} \neq 0$, hence $\mat f$ is invertible. Using that $\det{ \mat S_B} =1$ for every symplectic matrix~$\mat S_B$, equation \eqref{eq:sep_inequality_with_dets_pure} can be rewritten as
\begin{align}
 \det{\mat C} + \det{\mat{f}}^2\det{\mat{ C + S_B^{-1}f^{-1}[g]}} - 2 \det{\mat{f}} \det{\mat S} \nonumber\\
\leq 4 \det{\mat f}^2 \det{ \mat\gamma_{\mathrm{TMSS,r}} + (\zeroop \oplus\mat S_B^{-1}\mat f^{-1}[\mat g]) } + 1/4 .
\end{align}
Using that $\mat S_B^{-1} \mat f^-1[\mat g]$ is a symmetric matrix, this inequality can be evaluated (using computer algebra) and reduces to
\begin{equation}
  4 \det{\mat g} \geq \det{\mat f} ^2 +  2\det{\mat f} + 1 = (\det{\mat f} + 1) ^2 , \label{eq:channel_sep_condition_ineq_proof}
\end{equation}
which establishes \eqref{eq:channel_sep_condition_ineq}.

The generalization to the case of a Gaussian channel acting on one mode of a pure $N$-mode Gaussian state is straightforward: Any pure $N$-mode Gaussian input state can be transformed via a local $(N-1)$-mode unitary $\mat S_{A,N-1}$ and a local one-mode unitary $\mat S_B$  into a two-mode squeezed state on the mode $A_{N-1}$ and the mode $B$ as well as $N-2$ unrelated modes ($A_1,\ldots,A_{N-2}$) in the vacuum state with CM $\mat \gamma_{\mathrm{vac}} = \half \identityop$ \cite{botero03}:
\begin{align}
&\mat\sigma_{A_1,\ldots,A_{N-1},B}= \\
&\left( \mat S_{A,N-1}\oplus \mat S_B \right)
\left[ \left(\mat \gamma_{\mathrm{vac}}\right)_{A_1,\ldots,A_{N-2}} \oplus \left(\mat \gamma_{\text{TMSS},r}\right)_{A_{N-1},B} \right] . \nonumber
\end{align}
Again, the unitary $\mat S_{A,N-1}$ can be interchanged with the channel (which only acts on mode $B$) and undone. From here on, the proof for the case of an $N$-mode input state works exactly like the two-mode case since the modes in the vacuum state are irrelevant for the entanglement of the two modes in the two-mode squeezed state.
\end{proof}%
%
Summing up, \emph{qualitatively}, the entanglement of all pure Gaussian states evolves in the same way under a fixed one-sided Gaussian channel: Just as in the qubit case, the separability of the output state depends only on the channel, not on the initial state. The question whether or not it also evolves in the same way \emph{quantitatively}, hence whether or not a full Gaussian entanglement evolution equation similar to \eqref{eq:konrad_ent_evolution} exists, is still open. The answer to this question depends crucially on the choice of the right entanglement measure for a Gaussian version of \eqref{eq:konrad_ent_evolution}. Numerical calculations ruled out the existence of a Gaussian entanglement evolution equation for the two best-known entanglement measures in the context of Gaussian quantum information theory, the entanglement of formation and the logarithmic negativity \cite{hoelscher09} --- but, of course, this does not imply that no such equation can exist.

\vspace*{1em}

\emph{Addendum:} Right after completion of this work, Wang et al.~\cite{wang10} indeed published a Gaussian entanglement evolution equation using a different entanglement measure which is introduced in their article. This entanglement measure is a non-standard entanglement of formation based upon the shortest distance measure for pure states. Against the background of our numerical nogo results for a potential, quantitative entanglement evolution equation using the standard entanglement of formation \cite{hoelscher09}, the results of ref.~\cite{wang10} based on a non-standard entanglement of formation are very surprising.

\vspace*{1em}

\begin{acknowledgements}
We thank Raffaele Romano for very valuable discussions. 
J.H. acknowledges support through the BayBFG and the Studienstiftung des deutschen Volkes. 
P.v.L. acknowledges the DFG for financial support through the Emmy Noether programme.
\end{acknowledgements}

\end{document}